\documentclass[12pt]{article}
\usepackage{fullpage}
\usepackage{epsf}
\usepackage{epsfig}
\usepackage{amsthm}
\usepackage{amsfonts}
\usepackage{color}
\usepackage{amsmath}
\usepackage{natbib}

\begin{document}

\newcommand{\bm}[1]{\mbox{\boldmath $#1$}}
\newcommand{\mb}[1]{\mathbf{#1}}
\newcommand{\bE}[0]{\mathbb{E}}
\newcommand{\bP}[0]{\mathbb{P}}
\newcommand{\ve}[0]{\varepsilon}
\newcommand{\mN}[0]{\mathcal{N}}
\newcommand{\iidsim}[0]{\stackrel{\mathrm{iid}}{\sim}}
\newcommand{\NA}[0]{{\tt NA}}

\title{\vspace{-1cm} Optimization Under Unknown Constraints}
\author{Robert B.~Gramacy\\
  Statistical Laboratory\\
  University of Cambridge\\
  {\tt bobby@statslab.cam.ac.uk} \and
 Herbert K.H.~Lee\\
  Applied Math \& Statistics\\
  University of California, Santa Cruz\\
  {\tt herbie@ams.ucsc.edu}}
\date{}
\maketitle

\begin{abstract}
  Optimization of complex functions, such as the output of computer
  simulators, is a difficult task that has received much attention in
  the literature.  A less studied problem is that of optimization
  under unknown constraints, i.e., when the simulator must be invoked
  both to determine the typical real-valued response {\em and} to
  determine if a constraint has been violated, either for physical or
  policy reasons.  We develop a statistical approach based on Gaussian
  processes and Bayesian learning to {\em both} approximate the
  unknown function and estimate the probability of meeting the
  constraints. A new integrated improvement criterion is proposed to
  recognize that responses from inputs that violate the constraint may
  still be informative about the function, and thus could potentially
  be useful in the optimization.  The new criterion is illustrated on
  synthetic data, and on a motivating optimization problem from health
  care policy.

\end{abstract}

\section{Introduction}

A common optimization problem that arises in fields ranging from
applied engineering to public policy is to find $x^* =\mbox{arg}
\min_{x\in \mathcal{X}} f(x)$, subject to constraints: $x^* \in C$,
where we may only learn about the relationship between $x$ and
$f(x):\mathcal{X}\rightarrow \mathbb{R}$ and the constraint region $C$
through expensive evaluations of the noisy joint process
\begin{align}
  Z(x) &= f(x) + \ve,  \;\;\;\;\; \ve\sim \mN(0, \eta^2) \label{eq:proc} \\
  C(x) &= c(x + \ve_c) = \mathbb{I}_{\{x+\ve_c\in\, C\}} \in \{0,1\}. \nonumber
\end{align}
The real-valued noise variance, $\eta^2$, is unknown but may be zero,
and $\ve_c$ indicates that the constraint mapping may be random. In
particular, the constraint region $C\subset \mathcal{X}$ is
well-defined but often non-trivial.  Although it will typically be
deterministic ($\ve_c = 0$), this is not required by our treatment.
Finally, we suppose that observing the joint response $(Z,C)(x)$ is
expensive.  So we wish to keep the number of evaluations,
$(x_1,z_1,c_1), \dots (x_N, z_N,c_N)$, small.  One way to do this is
to build regression and classification models $f_N(x)$ for $f(x)$ and
$c_N(x)$ for $c(x)$ based on the data.  The surrogate surfaces may be
searched to find $x'$ yielding a small objective in expectation, and
satisfying the constraint with high probability.  We can then repeat
the process with $N+1$ points, including $(x', Z(x'), C(x'))$,
stopping when convergence in the location of $x^*$ is achieved, or
when resources are exhausted.

To shed light upon the difficulty in solving this problem, and to
thereby suggest possible points of attack, consider the following
simplification where the constraint region $C$ is known at the outset
(i.e., there is no need to estimate $c_N$).  In this case a sensible
approach is as follows.  Obtain realizations $z(x)$ of $Z(x)$ only for
$x \in C$ with the largest {\em expected improvement}
\citep[EI,][]{jones:schonlau:welch:1998} under $f_N$ [more on this in
Section \ref{sec:prevwork}] and proceed to construct $f_{N+1}$ by
adding the $(x, z(x))$ pair into the design.  This presumes that
evaluating $f(x)$ for $x \in \mathcal{X}\setminus C$ is a waste of
resources.  But this need not be so, since $Z(x)$, for {\em any} $x$,
contains information about $f$, and therefore about promising
location(s) for $x^*\in C$.  It could even be that $x' \notin C$ is
best at reducing the overall uncertainty in the location of $x^*\in
C$, through an improved new surrogate $f_{N+1}$.  When this is the
case [e.g., see Section \ref{sec:ieciill}] it makes sense to sample
$Z(x')$ for $x' \notin C$ despite the constraint violation.

Assessing when this odd maneuver is advantageous requires a more
global notion of improvement; EI cannot directly quantify the extent
to which $x'\notin C$ improves our information at $x \in C$.  Finally,
when $C$ is not known {\em a priori}, new evaluations $(x', z'=z(x'))$
provide information about both $f$ and $c$ through their surrogates
$f_N$ and $c_N$.  Thus incremental decisions toward solving the
constrained optimization problem must incorporate uncertainty from
both surrogates.  We propose a new integrated improvement statistic to
fit the bill.

The rest of the paper is outlined as follows.  In Section
\ref{sec:prevwork} we outline EI for (unconstrained) optimization and
the GP surrogate models upon which it is based.  In Section
\ref{sec:ciei} we develop the conditional and integrated expected
improvement statistic(s) for the case of known constraints, with an
illustration.  We extend the method to unknown constraints in Section
\ref{sec:uconst}, and demonstrate the resulting constrained
optimization algorithm on synthetic data.  In Section \ref{sec:rand}
we consider a motivating problem from health care policy research, and
conclude with some discussion and extensions in Section
\ref{sec:discuss}. Software implementing our methods, and the specific
code for our illustrative examples, is available in the {\tt plgp}
package \citep{plgp} for {\sf R} on CRAN.

\section{Previous Work}
\label{sec:prevwork}


\subsection{Surrogate Modeling}
\label{sec:surr}

The canonical choice of surrogate model for computer experiments is
the stationary Gaussian process
\citep[GP,][]{sack:welc:mitc:wynn:1989,ohagan99,sant:will:notz:2003},
which is one way of characterizing a zero mean random process where
the covariance between points is explicitly specified through the
function $C(x, x') = \sigma^2 K(x, x')$.  Let $Z_N=(z_1,\dots,z_N)^T$
be the vector of observed observed responses at the design points
$x_1, \dots, x_N$ collected (row-wise) in $X_N$.  Conditional on this
data $D_N= \{X_N, Z_N\}$, the (posterior) predictive distribution of
$Z(x)$ at a new point $x$ under the GP is normal with
\begin{align}
  \mbox{mean} && \hat{z}_N(x)  &= k_N^T K_N^{-1} Z_N, \label{eq:pvar} \\ 
\mbox{and variance} && \hat{\sigma}_N^2(x) & = \sigma^2 \nonumber
        [K(x,x) - k_N^T(x) K_N^{-1} k_N(x)], 
\end{align}
where $k_N^T(x)$ is the $N$-vector whose $i^{\mbox{\tiny th}}$
component is $K(x,x_i)$, and $K_N$ is the $N\times N$ matrix with
$i,j$ element $K(x_i, x_j)$.  These are sometimes called the {\em
  kriging equations}. Joint prediction at a collection of points $X$
is multivariate normal with mean vector $\hat{z}_N(X)$ and covariance
matrix $\hat{\Sigma}_N(X)$ which are defined by the straightforward
matrix extension of $k_N(X)$ and $K(X, X)$.  We follow
\cite{gra:lee:2008} in specifying that $K(\cdot,\cdot)$ have the form
$K(x, x'|g) = K^*(x, x') + \eta \delta_{x,x'}, \label{eq:cor} $ where
$\delta_{\cdot,\cdot}$ is the Kronecker delta function, and $K^*$ is a
{\em true} correlation function.  The $\eta$ term, referred to as the
{\em nugget}, is positive $(\eta>0)$ and provides a mechanism for
introducing measurement error into the stochastic
process---implementing $\eta^2>0$ in Eq.~(\ref{eq:proc})
\citep[][appendix]{gramacy:2005}.  It causes the predictive equations
(\ref{eq:pvar}) to smooth rather than interpolate the data $(X_N,
Z_N)$.
It is common to take $K^*(\cdot,\cdot)$ from a parametric family, such
as the separable Mat\'{e}rn or power families
\citep[e.g.,][]{abraham:1997}, which roughly model $K^*(\cdot,\cdot)$
as an inverse function of coordinate-wise Euclidean distance.  We
prefer the power family, which is standard for computer experiments.

\subsection{Optimization by Expected Improvement}
\label{sec:ei}

Conditional on a GP surrogate $f_N$, a step towards finding the
minimum may be based upon the {\em expected improvement} (EI)
statistic \citep{jones:schonlau:welch:1998}.  For a deterministic
function ($\eta=0$), the current minimum $f_{\min} =
\min\{z_1,\ldots,z_N\}$ is deterministic.  In this case, the
improvement is defined as $I(x) = \max\{f_{\min} - Z(x), 0\}$.  The
next location is chosen as
\begin{equation}
x' = \mbox{arg} \max_{x\in \mathcal{X}} \bE\{I(x)\},
\label{eq:ei}
\end{equation}
where the expectation is taken over $Z(x)\sim F_N(x)$, the predictive
distribution (\ref{eq:pvar}) implied by $f_N$ evaluated at $x$.
\cite{jones:schonlau:welch:1998} give an analytical expression for the
EI:
\begin{equation}
\bE\{I(x)\} = (f_{\min} - \hat{z}_N(x)) \Phi\left(
\frac{f_{\min} - \hat{z}_N(x)}{\hat{\sigma}_N(x)}\right)
+ \hat{\sigma}_N(x) \phi\left(
\frac{f_{\min} - \hat{z}_N(x)}{\hat{\sigma}_N(x)}\right).
\label{eq:eia}
\end{equation}
Basically, the EI is the cumulative distribution of the predictive
density that lies ``underneath'' $f_{\min}$. A relevant diagram
illustrating EI appears in Figure \ref{f:fmin} in Section
\ref{sec:fmin}.  \cite{jones:schonlau:welch:1998} also provide a
branch and bound algorithm for performing the maximization over
$\mathcal{X}$ to find $x'$.  Once $x'$ is chosen it is added into the
design as $(x_{N+1}, z_{N+1}) = (x', f(x'))$ and the procedure repeats
with $f_{N+1}$.  \cite{jones:schonlau:welch:1998} use maximum
likelihood inference to set the parameters for $f_N$, i.e., $d$ only
since $\eta=0$, and call the resulting iterative procedure the {\em
  efficient global optimization} (EGO) algorithm.  The above choice of
$f_{\min}$ is sensible but somewhat arbitrary.  Another reasonable
choice that we promote in this paper is $f_{\min} = \min
\hat{z}_N(x)$, the minimum of the (posterior) mean predictive
surface.

The situation is more complicated for noisy responses. We must then
estimate the nugget, $\eta$, and extend the
\cite{jones:schonlau:welch:1998} definition of $f_{\min}$ to be a
random variable: the first order statistic of $Z_1,\dots, Z_N$.
Calculating the EI would thus require integrating over $f_{\min}$ in
Eq.~(\ref{eq:ei}).  This breaks the analytical tractability of EGO
algorithm, however one can always proceed by Monte Carlo methods.
Once in the Monte Carlo framework, extensions abound.  For example, it
is trivial to take a Bayesian approach and thereby factor parameter
uncertainty into the EI calculation.  Conditional on the parameters
however, choosing $f_{\min}\min \hat{z}_N(x)$ is still deterministic.
So this choice allows an analytical approach to proceed when
point-estimates (i.e., MLEs) of parameters are used, or it leads to a
more efficient Monte Carlo algorithm when sampling from the
Bayesian posterior.  The downside of the Monte Carlo approach, whether
taken for Bayesian or $f_{\min}$ considerations, is that the branch
and bound algorithm for determining $x'$ in Eq.~(\ref{eq:ei}) is no
longer available.  However, proceeding with a discrete set of
space-filling candidates, and leveraging direct optimization methods
in tandem, has proved fruitful \citep{tadd:lee:gray:grif:2009}.

\subsection{Towards Constrained Optimization}
\label{sec:const}


Ours in not the first attempt at tackling the constrained optimization
problem via surrogate modeling.  \cite{scho:welc:jone:1998} consider
deterministic responses ($\eta=0$) where the known constraint region
can be written as $a_k \leq c_k(x) \leq b_k$, for $k=1,\dots,K$.  They
then treat the $c_k(x)$ as additional response variables that co-vary
with $f(x)$.  This breaks the analytical tractability of the EI
calculation.  Upon assuming that the $K+1$ responses are independent
the calculation is again tractable, otherwise a Monte Carlo approach
is needed.
We are not aware of any previous literature addressing our more
general problem: where the function $f$ may not be deterministic, and
when there are unknown constraints of arbitrary form.  Even in simpler
settings, like the one above, it may be advantageous to sample outside
the constraint region.  This requires a new improvement
statistic---one that weighs the overall expected improvement of the
next sequentially chosen design point in aggregate.

\section{Integrated Expected Conditional Improvement}
\label{sec:ciei}

Here we generalize the EI framework to accommodate the drawbacks
outlined above.  To start with, we assume that constraints are
deterministic, and known (with trivial computation) in advance.
Section \ref{sec:uconst} provides extensions for unknown constraints.

Define the {\em conditional improvement} as
\begin{equation}
I(y| x) = \max\{f_{\min} - Z(y|x), 0\}, \label{eq:condi}
\end{equation}
where $Z(y|x) \sim F_N(y|x)$, which is the predictive distribution of
the response $Z(y)$ at a {\em reference} input location $y$ under the
surrogate model $f_N$ given that the {\em candidate} location $x$ is
added into the design.  We do not use an $N+1$ subscript for the
posterior predictive distribution because the realization of the
response $z(x)$ is not yet available.  

The {\em expected conditional improvement} (ECI) at the reference
point $y$ is then $\bE\{I(y|x)\}$. Here the expectation is over all of
the random quantities: the distribution of $Z(y|x)$, and perhaps of
$f_{\min}$ depending upon how it is defined.  The ECI may be evaluated
at all pairs of inputs $(x,y) \in \mathcal{X}$. The potential to
generalize EI, which accounts for improvement at the point $x$ alone,
comes by integrating over the choices for $y$.  Let $g(y)$ denote a
density over $y \in \mathcal{X}$ which may be uniform in a bounded
region.  Then the {\em integrated expected conditional improvement}
(IECI) is defined as
\begin{equation}
\bE_g\{I(x)\} = -\int_{\mathcal{X}}\bE\{I(y|x)\}g(y) \,dy.
\label{eq:iei}
\end{equation}
This suggests using $ x' = \mbox{arg} \max_{x\in \mathcal{X}}
\bE_g\{I(x)\}$ as the next adaptively sampled point.  As long as
$\bE\{I(y|x)\} \leq \bE\{I(y)\}$ for all $x\in \mathcal{X}$, this
statistic (\ref{eq:iei}) is defensible.  Defining $f_{\min}$ carefully
[see Section \ref{sec:fmin}] ensures that this {\em
  monotonicity condition} holds.

The negation in Eq.~(\ref{eq:iei}) keeps IECI in line with the
convention of maximizing, i.e., of preferring large EI statistics over
small ones. To explain, consider how $I(y|x)$ ``looks ahead''.  We
wish to measure an improvement at $x$, but in a roundabout way we
assess that improvement at a reference point $y$ instead, supposing
$x$ has been added into the design.  If $y$ still has high improvement
potential after $x$ has been added in, then $x$ must not have had much
influence on the improvement at $y$.  If $x$ is influential at $y$,
then the improvement at $y$ should be small after $x$ is added in, not
large.

We can alternatively define IECI as the expected reduction in
improvement at the reference location, $y$, when $x$ is added into the
design:
\begin{equation}
\bE_g\{I(x)\}  = \int_{\mathcal{X}} (\bE\{I(y)\} - \bE\{I(y|x)\}) g(y)
\,dy, \label{eq:reduce}
\end{equation}
which is guaranteed to be positive under our monotonicity assumption.
We would then take the $x'$ which gave the largest reduction.  But
clearly this is within an additive constant (the weighted-average EI
over $g(y)$) of the definition given in Eq.~(\ref{eq:iei}), and is
thus equivalent.

The integrated approach allows constraints to be handled through
$g(y)$.  E.g., $g(y)$ can be uniform for $y\in C$ and zero otherwise.
Or, [as we discuss in Section \ref{sec:uconst}] it can give higher
weight to $y$ with a greater chance of satisfying the constraint.
When there are no constraints, choosing $g(y)$ uniform on $y \in
\mathcal{X}$ yields an aggregated statistic that will offer a more
global search, compared to EI, in a manner similar to how the expected
reduction in variance generalizes the predictive variance for
sequential design by active learning
\citep{seo00,gra:lee:2009}.  

\subsection{Expected Conditional Improvement}
\label{sec:eci}

The key ingredient in calculating the ECI is an assumption about how
$Z(y|x)$ behaves relative to $Z(y)$.  Let $F_N(y|x)$ denote the
distribution of
$Z(y|x)$.  
Overloading the notation somewhat, let $f_N(z(x))$ denote the density
of $Z(x)$ under $F_N$, and likewise $f_N(z(y)|x)$ for $Z(y|x)$. By the
law of total probability,
\begin{align}
f_N(z(y)|x) &= \int f_N(z(y), z(x)|x) \;dz(x) \label{eq:integral} \\
&= \int f_{N+1}(z(y) | x, z(x)) f_N(z(x)) \; dz(x), \nonumber
\end{align}
where $f_{N+1}(z(y) | x, z(x))$ is the predictive density of $Z(y)$
when the design matrix and response vector are augmented by $(x,
z(x))$.  Note that the above expressions involving $z(y)$ have an
implicit conditioning upon $y$.  For an arbitrary surrogate, computing
the integral in Eq.~(\ref{eq:integral}) analytically would present a
serious challenge.  However, under a GP surrogate it is trivial since
$F_N$ and $F_{N+1}$ are both (univariate) normal distributions
(\ref{eq:pvar}), and a convolution of normals is also normal.
Trivially, the mean and variance of the (normal) predictive density
$f_{N+1}(z(y) | x, z(x))$ is unchanged after integrating out $Z(x)$
since the GP is not dynamic, so there is no update from $f_N$ without
observing $z(x_{N+1})$.

But at the same time, the predictive variance (\ref{eq:pvar}) does not
depend upon the responses, $Z_N$ or $z(x)$ via $Z_{N+1}$.  So we can
deduce what the variance of the predictive density $f_{N+1}(z(y) | x,
z(x))$ {\em will be} once $z(x)$ arrives. We will have
$\hat{\sigma}_{N+1}^2(y|x, z(x)) = \hat{\sigma}_{N+1}^2(y|x)$ under
the {\em assumption} that the evidence in $z(x)$ does not
update/change parameters of the GP (which it can't if it is not
observed!).  Now, $\hat{\sigma}_{N+1}^2(y|x, z(x))$ depends upon
$K_{N+1}^{-1}(x)$ whose $N+1^{\mathrm{st}}$ row and column are
populated with $K(x_i, x)$ for $i=1,\dots,N$ and with $K(x,x)$
appearing in the bottom right-hand corner.  So $K_{N+1}^{-1}(x)$ can
then be obtained in terms of $K_N^{-1}$ via partitioned inverse
equations.  If
\begin{align*}
K_{N+1}(x) &= \begin{bmatrix}
K_N & k_N(x) \\
k_N^T(x) \hspace{-.1cm} & K(x,x)
\end{bmatrix}, \;\;\;\;\;  \mbox{then}   \\
K_{N+1}^{-1}(x) &= \begin{bmatrix}
[K_N^{-1} + g(x) g^T(x) \mu^{-1}(x)] & g(x) \\
g^T(x) & \mu(x)
\end{bmatrix},
\end{align*}
where $g(x) = -\mu(x) K_{N}^{-1} k_N(x)$ and $\mu^{-1}(x) = K(x,x) -
k_N^T(x) K_N^{-1} k_N(x)$.  This saves us from performing any
additional $O(N^3)$ matrix operations.  So $\hat{\sigma}_{N+1}^2(y|x)
= \sigma^2 [K(y,y) - k_{N+1}^T(x; y) K_{N+1}^{-1}(x) k_{N+1}(x;
y)]$ where $k_{N+1}^T(x; y)$ is an $(N+1)$-vector whose first $N$
entries are identical to $k_N(y)$ and with an $N+1^{\mathrm{st}}$
entry of $K(y,x)$.  The amount by which
$\hat{\sigma}_{N+1}^2(y|x,z(x))$ is reduced compared to
$\hat{\sigma}_{N}^2(y)$ is then readily available.  Let $G(x)\equiv g(x)
g^T(x)$.  Then,
\begin{align}
\hat{\sigma}_{N+1}^2(y|x) =  \label{eq:xyvar}
\hat{\sigma}_{N}^2(y) - \sigma^2 [&k_N^T(y) G(x) \mu^{-1} k_N(y) \\
&+ 2 k_N^T(y) g(x) K(x,y) + K(x,y)^2\mu ]. \nonumber
\end{align}
So we can see that the {\em deduced} predictive variance at $y$ will
be reduced when $z(x)$ is observed by an amount that depends upon how
far apart $y$ and $x$ are.  This is not only sensible, but will also
be helpful for determining the influence of $x$ in improvement
calculations.

To sum up, we propose to {\em define} $F_N(y|x)$, for the purposes of
sequential design, to be a normal distribution with (true) mean
$\hat{z}_N(y|x) = \hat{z}_N(y)$ and deduced variance
$\hat{\sigma}_N^2(y|x) \equiv \hat{\sigma}_{N+1}^2(y|x, z(x)) =
\hat{\sigma}_{N+1}^2(y|x)$ as given in Eq.~(\ref{eq:xyvar}), above.
As with the kriging equations (\ref{eq:pvar}), joint sampling for a
collection of ($M$) reference inputs $Y_M$ is possible via the
appropriate matrix extensions to $k_N(Y_M)$ and $K(Y_M,Y_M)$ in order
to derive $\hat{z}_N(Y_M|x)$ and $\hat{\Sigma}_N(Y_M|x)$.

Now, with an appropriate definition of a deterministic $f_{\min}$, the
same analytic expression for the EI from Section \ref{sec:prevwork}
can be extended to the ECI:
\begin{align}
\bE&\{I(y|x)\} =  \label{eq:eci} \\
& (f_{\min} - \hat{z}_N(y|x)) \Phi\!\left( 
\frac{f_{\min} - \hat{z}_N(y|x)}{\hat{\sigma}_N(y|x)}\right)
+ \hat{\sigma}_N(y|x) \phi\!\left(
\frac{f_{\min} - \hat{z}_N(y|x)}{\hat{\sigma}_N(y|x)}\right).
\nonumber
\end{align}
If we cared only about the ECI (without integration (\ref{eq:iei})),
the branch and bound algorithm given by
\cite{jones:schonlau:welch:1998} would apply leading to a {\em
  conditional} EGO algorithm.

\subsubsection{Choosing $f_{\min}$}
\label{sec:fmin}

\begin{figure}[ht!]
\centering
\vspace{0.25cm}
\input{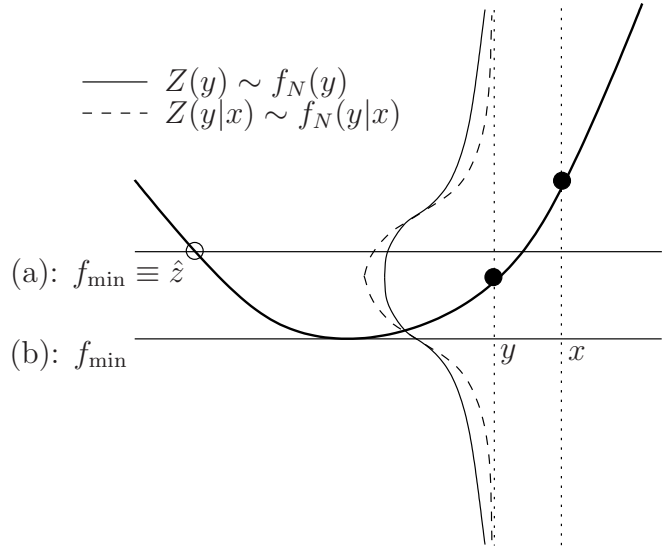}
\vspace{0.25cm}
\caption{Illustrating how the choice of $f_{\min}$ influences
  the ECI.  The solid curve represents the mean-predictive 
  $\bE\{Z(\cdot)\}$.  The densities
  of $Z(y)$ and $Z(y|x)$ are shown as solid and dashed
  ``bell-curves'', respectively.  In (a) $f_{\min}$ is taken
  to be the mean predictive at the $N$ input locations whereas in (b)
  it is taken to be the minimum of predictive-mean surface.  The respective
  improvements are the areas of the densities underneath
  $f_{\min}$.}
\label{f:fmin}
\end{figure}

Figure \ref{f:fmin} illustrates how a deterministic choice of
$f_{\min}$ can influence the ECI.  Consider two cases ((a) and (b)),
which pertain to the choices for $f_{\min}$ introduced in Section
\ref{sec:ei} (represented by horizontal lines): (a) uses only the
observed locations and (b) uses the whole predictive curve.  We will
return to details of these choices shortly.  In the figure, the solid
parabolic curve represents the predictive mean surface
$\bE\{Z(\cdot)\}$.  The EI is the area of the predictive density drawn
as a solid line, plotted vertically and centered at $\hat{z}(y)$,
which lies underneath the horizontal line(s), representing choices of
$f_{\min}$.
The ECI is likewise the area of the predictive density drawn as a
dashed line lying below the horizontal line(s).  This dashed density
has the same mean/mode as the solid one, but it is more sharply peaked
by the influence of $x$.  If we suppose that the densities, drawn as
bell-curves in the figure, are symmetric (as they are for a GP), then
it is clear that the relationship between ECI and EI depends upon
$f_{\min}$.  As the dashed line is more peaked, the left-tail
cumulative distributions have the property that $F_N(f_{\min}|x) \geq
F_N(f_{\min})$ for all $f_{\min} \geq \bE\{Z(y|x)\} = \bE\{Z(y)\}$, to
which choice (a) for $f_{\min}$ corresponds.  Therefore $\bE\{I(y|x)\}
\geq \bE\{I(y)\}$ in this case, violating our desired monotonicity
property.  But for choice (b) the ECI represents a reduction compared
to the EI, since $f_{\min} \leq \bE\{Z(y|x)\}$, thus satisfying the
monotonicity property.

Case (a) in Figure \ref{f:fmin} is meant to represent taking
$f_{\min}=\min\{z_1,\dots,z_N\}$, deterministically.  It may similarly
represent the minimum of the mean-predictive at the $X_N$ locations,
which would coincide with the minimum of the $Z_N$ values in the
no-noise ($\eta = 0$) case.  In the noisy case ($\eta > 0$) $f_{\min}$
in Eq.~(\ref{eq:condi}) is a random variable whose distribution can be
approximated by simulation from $F_N$.  But this extra computational
effort would be in vain because the monotonicity property is not
guaranteed.  Case (b) corresponds to taking $f_{\min} = \min
\bE\{Z(\cdot)\}$, the minimum of the posterior
mean-predictive---another deterministic choice.  In this case it is
clear that $f_{\min}$ will always cut through the density of $Z(y|x)$
at or below its mean/mode $\bE\{Z(y|x)\} = \bE\{Z(y)\}$ and ensure
that the monotonicity property is satisfied.  Accordingly, we shall
use this choice throughout the remainder of the paper.

\subsubsection{A Monte Carlo Approach for Calculating the ECI}
\label{sec:ecimc}

\begin{figure}[ht!]
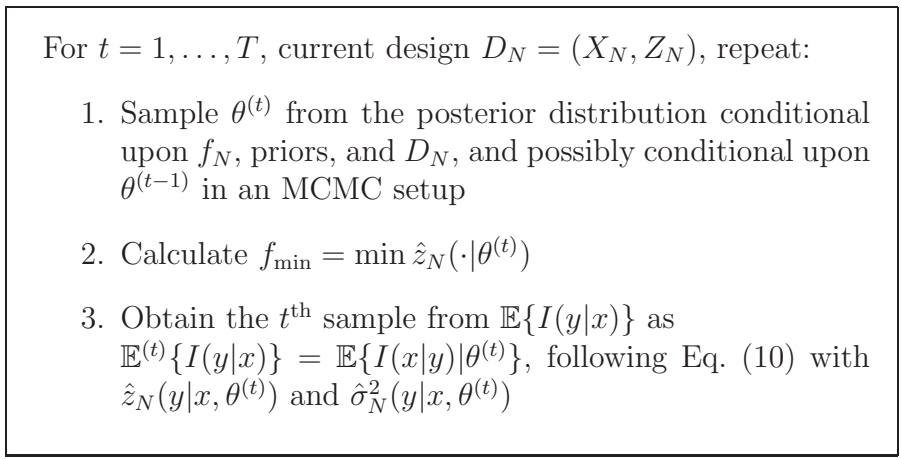

   \centering \fbox{ \hspace{0.1cm} \begin{minipage}{11cm}
      \vspace{0.25cm} For $t=1,\ldots,T$, current design $D_N = (X_N,
      Z_N)$, 
     repeat:
\begin{enumerate}
\item Sample $\theta^{(t)}$ from the posterior distribution
  conditional upon $f_N$, priors, and $D_N$, and possibly conditional
  upon $\theta^{(t-1)}$ in an MCMC setup
 \item Calculate $f_{\min} = \min \hat{z}_N(\cdot|\theta^{(t)})$
\item Obtain the $t^{\mathrm{th}}$ sample from $\bE\{I(y|x)\}$ as \\
$
\bE^{(t)}\{I(y|x)\} = \bE\{I(x|y)|\theta^{(t)}\}
$, 
following Eq.~(\ref{eq:eci}) with $\hat{z}_N(y|x,\theta^{(t)})$ and
$\hat{\sigma}^2_N(y|x, \theta^{(t)})$
\end{enumerate}
\vspace{0.1cm}
\end{minipage}\hspace{0.25cm}}
\vspace{0.25cm}
\caption{Monte Carlo approximation of the ECI statistic.}
\label{alg:MCeci}
\end{figure}

The following procedure may be used to obtain samples of the ECI via
the GP surrogate posterior predictive $f_N$, taking full account of
uncertainty in the parameters $\theta = (\sigma^2,d,\eta)$.  The
procedure is borne out via Monte Carlo sampling for $\theta$ in Figure
\ref{alg:MCeci}.
If $\theta$ is considered known, or has been estimated offline, e.g.,
via maximum likelihood, then we may skip the loop (and Step 1),
taking $T=1$ with $\theta^{(1)} = \theta$.  In either case, an
estimate of the ECI is obtained by ergodic averaging:
\begin{equation}
\bE\{I(y|x)\} \approx \frac{1}{T} \sum_{t=1}^T \bE^{(t)}\{I(y|x) \}.
\label{eq:eecigp}
\end{equation}

\subsection{Integrated Expected Conditional Improvement Algorithm}
\label{sec:iei}

Calculating the IECI (\ref{eq:iei}) from the ECI requires integrating
over $y \in \mathcal{X}$ according to $g(y)$, which may be uniform in a
bounded (constraint) region.
It will not generally be possible to integrate analytically, so we
propose to augment the Monte Carlo procedure from Section~\ref{sec:ecimc}.
Given a large number of sampled reference locations $Y_M \equiv
y^{(1)},\dots,y^{(M)} \iidsim g$, the IECI may be approximated
with $T$ Monte Carlo samples from the ECI as follows.
\begin{align}
\bE_g\{I(x)\} &\approx  
-\frac{1}{MT} \sum_{m=1}^M \sum_{t=1}^T \bE^{(t)}\{I(y^{(m)}|x)\}
\label{eq:biia}
\end{align}
When the parameters $\theta$ are known, $T=1$ as before.  With larger
$M$ (and $T$) we obtain an improved approximation, and in the limit we
have equality.  In the case where $g$ is uniform over a convex region,
a grid or maximum entropy design may be preferred \citep[][Section
6.2.1]{sant:will:notz:2003}.  When the marginals of $g$ are known, a
Latin Hypercube Design \citep[LHD,][Section
5.2.2]{sant:will:notz:2003} may be more generally appropriate.

If we choose (or are required) to work with a size $M$ grid, design,
or LHD of reference locations $y\in \mathcal{X}$, we may view $g$ as
discrete and finite measure.  An alternate approach in this case is to
forgo (re-)sampling from $g$ and compute a weighted average instead:
\begin{equation}
  \bE_g\{I(x)\} \approx -\frac{1}{T} \sum_{t=1}^T \frac{1}{M} \sum_{m=1}^M
  \bE^{(t)}\{I(y^{(m)}|x)\} g(y^{(m)}). \label{eq:bmciia}
\end{equation}
This has the disadvantage that the ECI may be evaluated at many
reference locations $y^{(m)}$ with low (or zero) probability under
$g$. But it has the advantage of an implementation that is easily
adapted to the unknown constraint situations described
shortly.  

\subsection{Illustrating IECI}
\label{sec:ieciill}

\begin{figure}[ht!]
\centering
\vspace{0.3cm}
\includegraphics[scale=0.5,trim=15 0 0 0]{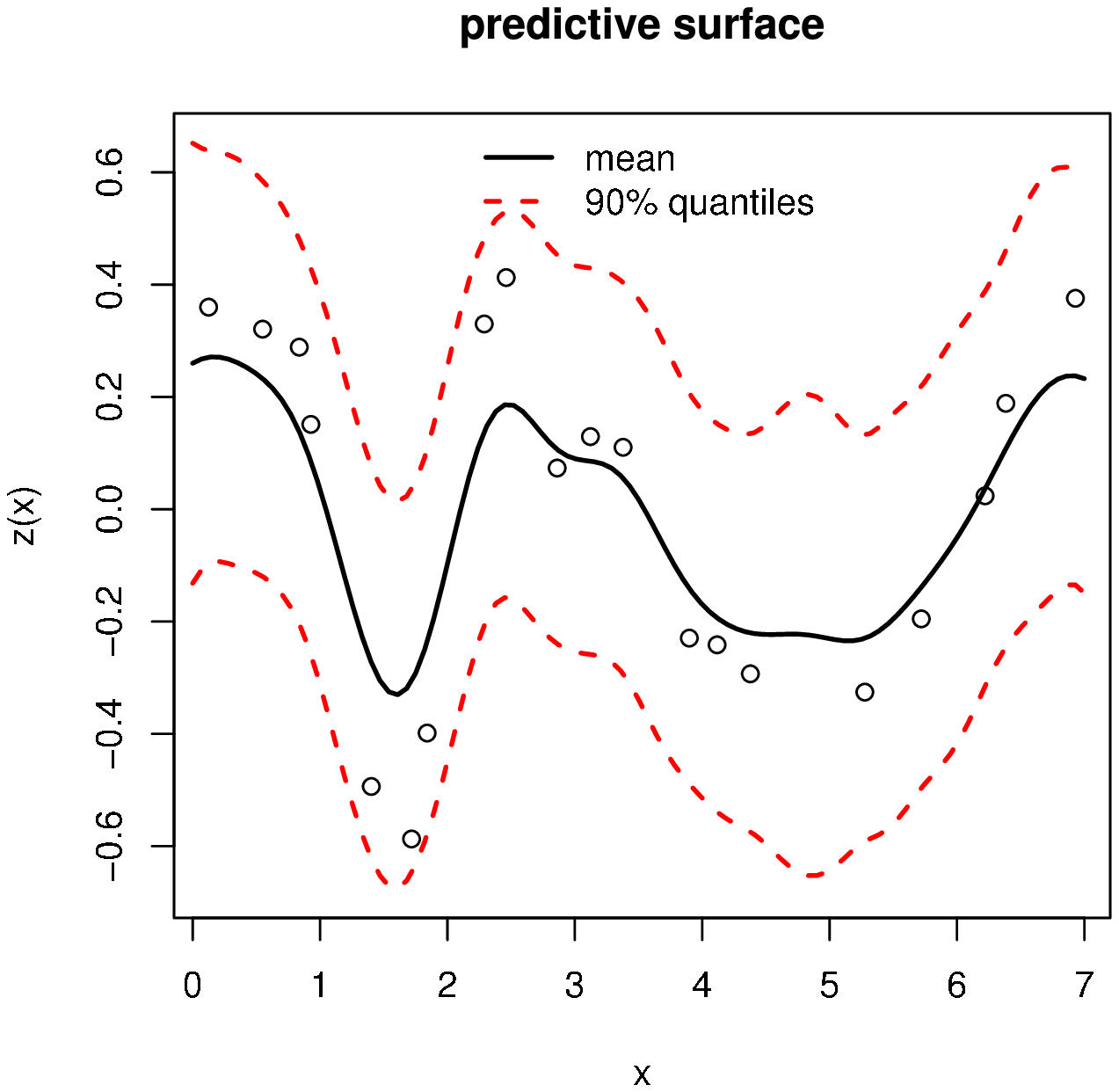}
\hspace{0.2cm}
\includegraphics[scale=0.5, trim=0 0 10 0]{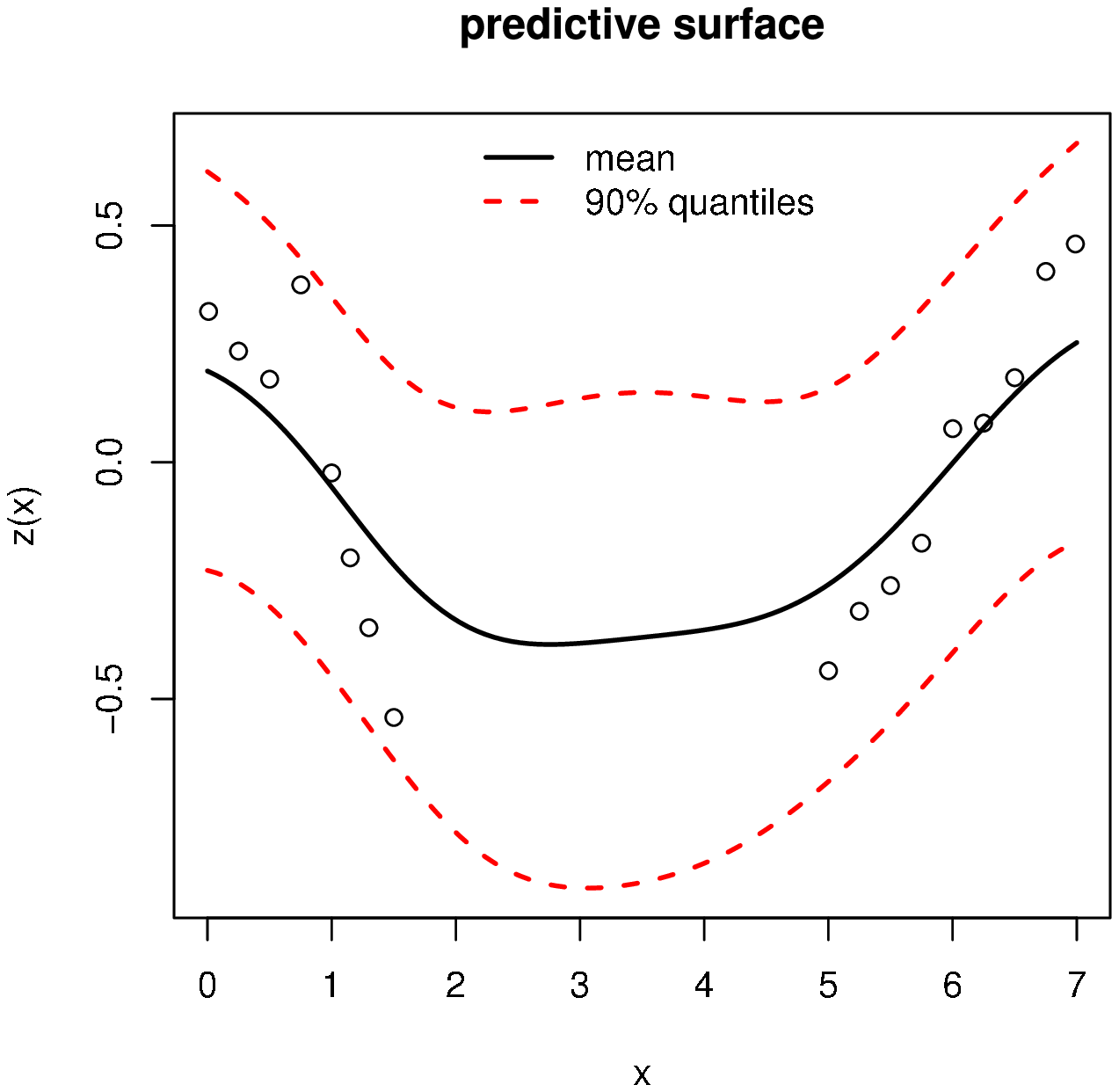}\\
\vspace{0.25cm}
\includegraphics[scale=0.5, trim=15 10 0 0]{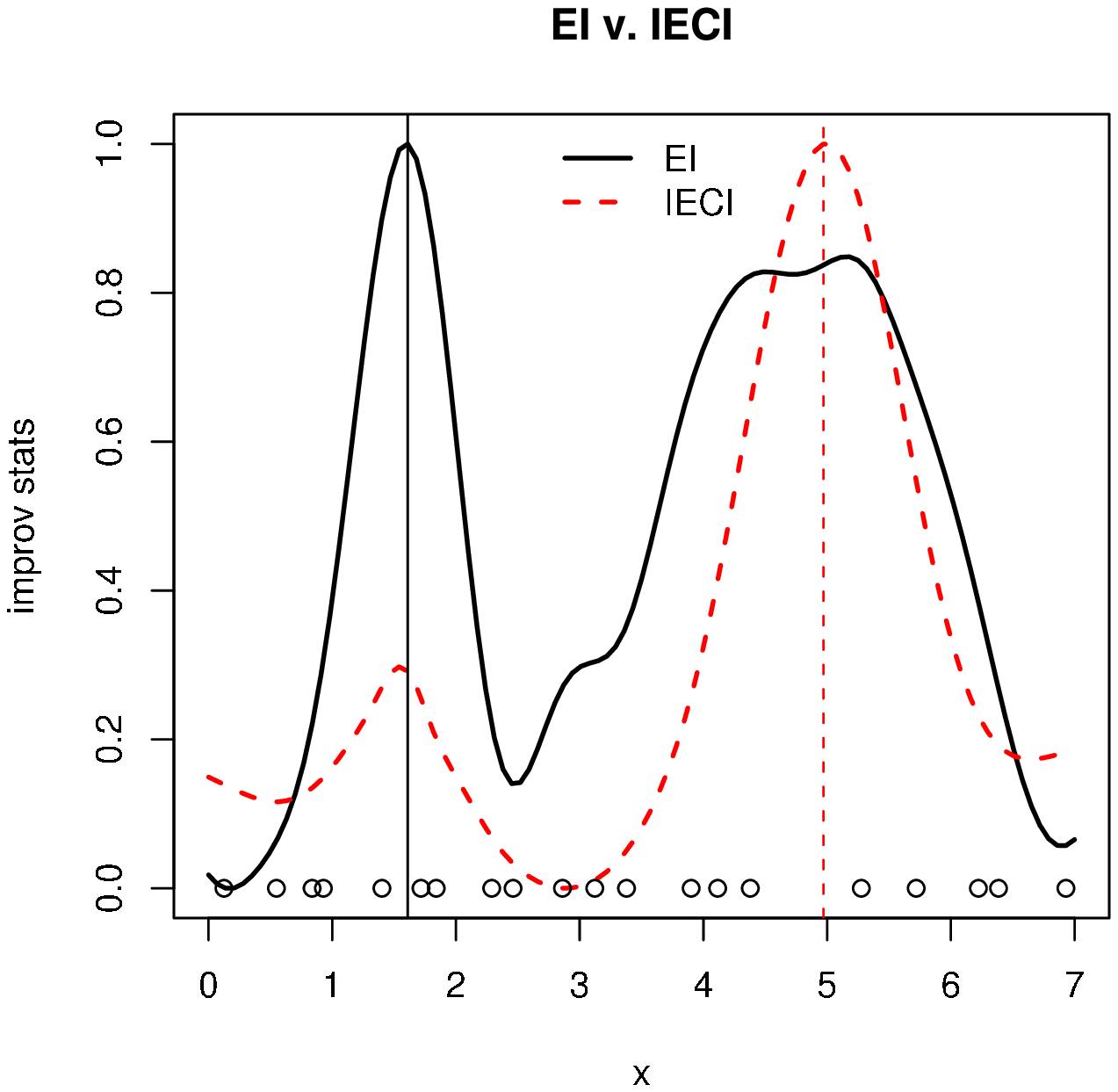}
\hspace{0.2cm}
\includegraphics[scale=0.5, trim=0 10 10 0]{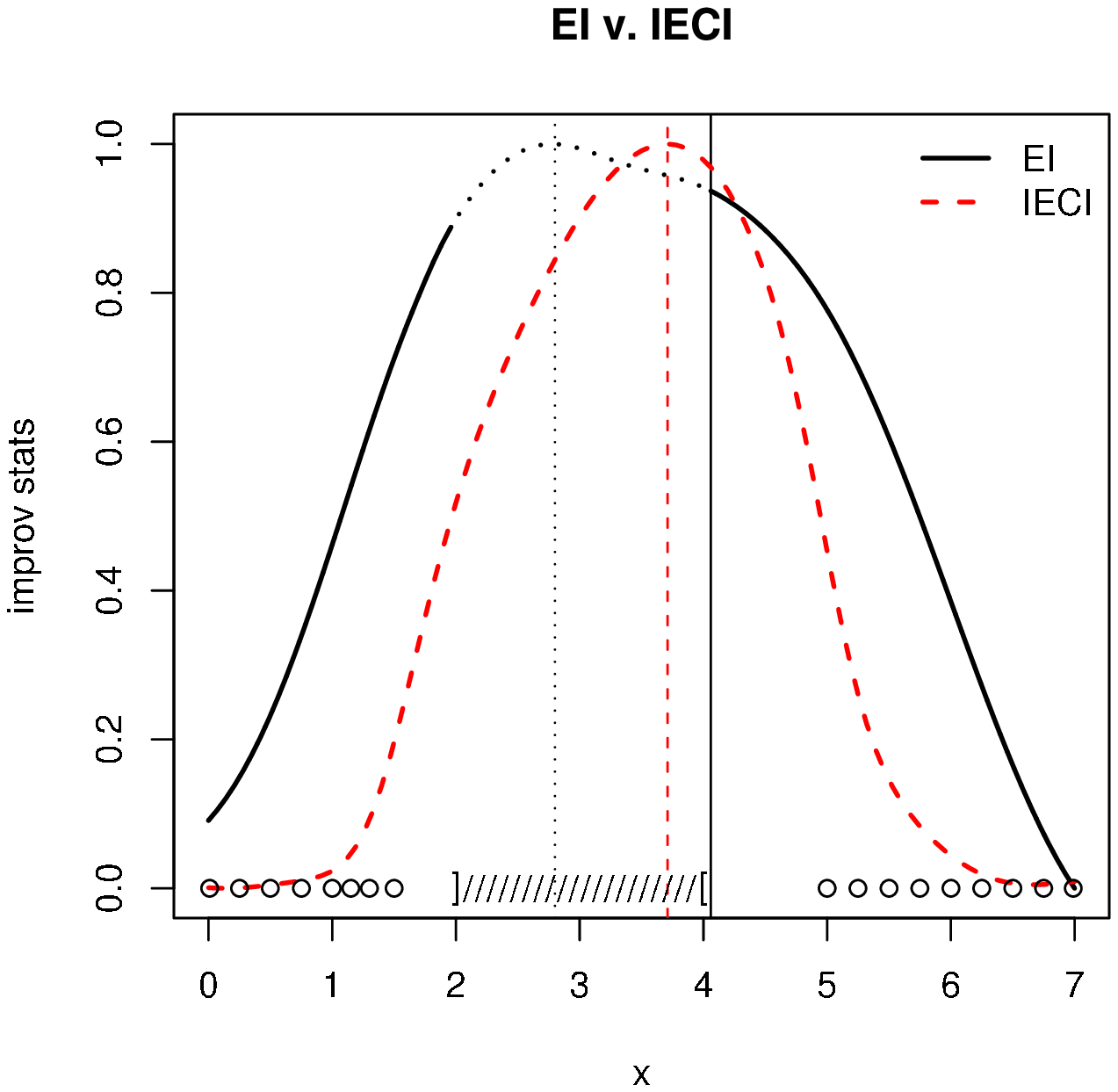}
\caption{Comparing EI and IECI.  The {\em top} panels show the design
  and posterior predictive surface.  The {\em bottom} panels show EI
  and IECI statistics for the corresponding surfaces above.  In the
  case of constrained optimization, in the {\em right} panels, the
  constraint violation region $C^c$ is shown with slashes.}
\label{f:ieci}
\end{figure}

To illustrate IECI consider the following process $ \bE\{Z(x)\} = f(x)
= \sin(x) + 2.55\phi_{0.45}(x-3)$, observed for $x\in[0,7]$.  As a
mixture of a sinusoid and normal density function (with $\mu=3$ and
$\sigma=0.45$) it has two local minima in this region.  To make things
interesting, realizations of the process are observed with
i.i.d.~noise so that Var$\{Z(x)\}=0.15^2$.  The {\em top-left} panel
of Figure \ref{f:ieci} shows one random realization of this process at
LHD inputs.
The predictive mean and 90\% interval obtained by sampling from the
posterior under the GP is also shown.  A visual inspection of the
surface(s) reveals that, indeed, there are two local minima.

Below that panel, on the {\em bottom-left}, the EI (solid black) and
IECI (dashed red) surfaces are plotted, normalized to appear on the
same $[0,1]$ scale.  As a further visual aid, the design $X_N$ is also
shown, and the vertical lines crossing the $x$-axis intersect with the
curves at their maxima.  We took a uniformly spaced set of 100
candidate locations in $[0,7]$, our $\mathcal{X}$, and calculated the
EI and IECI at $x\in \mathcal{X}$.  Likewise, we took the same $M=100$
points as reference locations $Y_M = \mathcal{X}$ for the IECI
calculations via Eq.~(\ref{eq:biia}).  EI recommends taking a sample
from the left local minima, although the relative heights of the two
disparate regions of highest EI betrays that this decision is indeed a
``close call''.  In contrast, IECI suggests taking the next sample
from the right local minima, and with much greater decisiveness.  The
lower concentration of samples nearby this right-minima lead to higher
variance in that region which may be pooled by the more
globally-scoped IECI.

The {\em right-hand} panels in Figure \ref{f:ieci} show a similar
sequence of plots in the presence of a known constraint $C = [0,2]
\cup [4,7]$.  To illustrate EI and IECI in this scenario, consider the
random realization and corresponding posterior predictive surface in
the {\em top-right} panel.  Here the $X_N$ design locations all
reside inside $C$.  The {\em bottom-right} panel shows the EI
statistic over the entire (discrete) range for $x\in \mathcal{X}$, as
above.  Those parts of the EI curve corresponding to inputs which
violate the constraint are dotted.  The EI is maximized outside of the
constraint region near $x=2.75$, with the maximal value inside $C$ at
the $x=4$ boundary.  The IECI statistic is also shown over the entire
range, but the $y^{(m)}$ locations are restricted to $C$.  I.e., $Y_M
= \mathcal{X} \cap C$.  This is so that we may consider the extent to
which every location $x\in\mathcal{X}$ reduces the average conditional
improvement $y\in C$.  Observe that the maximal IECI point is
$x=3.75$.  This point gives the greatest reduction in improvement
averaged over the constraint region, even though it does not, itself,
satisfy the constraint.

\section{Dealing with Unknown Constraints}
\label{sec:uconst}

Here we extend the IECI to unknown constraints.  Much of the necessary
scaffolding has already been built into the IECI via $g(y)$, e.g.,
$g(y) = \bP(C(y) = 1)$.  It remains for us to flesh out the Monte
Carlo by incorporating the surrogate $c_N$ for $C(y)$.  We extend the
parameter vector $\theta$ to contain parameters for both surrogates:
$\theta = (\theta_f, \theta_c)$; and the data to include the
class/constraint labels: $D_N = (X_N, Z_N, C_N)$.  Inference for
unknown $\theta|D_N$ is via samples from the joint posterior.  An
appropriate choice of $c_N$ is discussed in Section~\ref{sec:plgp}.

For now, overload the generic classification surrogate notation to let
$c_N(y^{(m)}|\theta_c^{(t)})$ denote the probability input $y^{(m)}$
satisfies the constraint given parameters $\theta_c^{(t)}$.  Then,
\begin{equation}
 \bE_c\{I(x)\} \approx -\frac{1}{T} \sum_{t=1}^T \frac{1}{M} \sum_{m=1}^M
  \bE^{(t)}\{I(y^{(m)}|x)\} \cdot c_N(y^{(m)}|\theta_c^{(t)}).
 \label{eq:bumciia}
\end{equation}
Note that in $\bE^{(t)}\{I(y^{(m)}|x)\}$ there is an implicit
dependence upon $\theta_f^{(t)}$, unless these parameters are taken as
known.  In that case we may drop the $(t)$ superscript from the ECI
expression in Eq.~(\ref{eq:bumciia}), and re-arrange the order of
summation to avoid unnecessarily re-calculating the ECI for each $t$.
Observe that Eq.~(\ref{eq:bumciia}) extends Eq.~(\ref{eq:bmciia})
rather than (\ref{eq:biia}).  Sampling from the surrogate $g_N$,
rather than simply evaluating the related quantity $c_N$, would not
generally be straightforward, and so we prefer to work with 
design-based candidates $y \in \mathcal{X}$.

\subsection{An Appropriate Constraint Surrogate, and Sequential Inference}
\label{sec:plgp}

An appropriate partner to the canonical GP (regression) surrogate
$f_N$ for $f$ is a classification GP (CGP) surrogate $c_N$ for $c$.
For details on CGP specification and corresponding Monte Carlo
inference based on MCMC, see \cite{neal:1998}.  As in the regression
case, the CGP model is highly flexible and competitive with, or better
than, the most modern models for non-parametric classification.
However, batch inference methods based on MCMC are at odds with the
sequential nature of the design strategy.  Except to guide the
initialization of the new Markov chain, it is not clear how fits from
earlier iterations may re-used in search of the next design point. So
after each sequential design step the MCMC must be re-started and
iterated until convergence.  The result is a slow algorithm.

So instead of taking the traditional, established, MCMC approach to
C/GP inference we follow a new sequential Monte Carlo
(SMC) approach outlined by \cite{gramacy:polson:2010}.  They show how
GP and CGP models can be implemented in an online setting, by
quickly updating a discrete approximation to the posterior via
{\em particle learning} \citep{carvalho:etal:2008}.  This approach
leads to fast online---and in some cases statistically superior
(i.e., lower MC error)---posterior summaries compared to MCMC.
\cite{gramacy:polson:2010} go on to describe to how EI for
optimization and entropy based boundary exploration for classification
can proceed efficiently with particles.  This is easy to extend to
IECI by coupling the regression and classification models ($f_N$ and
$c_N$) via the Monte Carlo approximations described earlier in this
paper.

\subsection{Illustrations and Examples}
\label{sec:cill}

We provide two synthetic data examples where the constraint region is
unknown.  In both cases we take the candidate and reference locations
(identically: $Y_m = \mathcal{X}$) as a LHD randomly generated at the
beginning of each round and then augmented with an {\em oracle point}
\citep{tadd:lee:gray:grif:2009}.  We follow \cite{gramacy:taddy:2010}
in taking the oracle point as the local maximum obtained via numerical
non-derivative minimization initialized at the last added design point
and searched over the MAP predictive surface inferred in the previous
round.  An implementation via particles is described by
\citep{gramacy:polson:2010}.

The objective function and constraint region for the first example was
presented in Section \ref{sec:ieciill}.  We initialize the
optimization with a size 20 LHD, and then collect 60 points by IECI
with 100 fresh candidates/reference locations as described above.
\begin{figure}[ht!]
\centering
\vspace{0.3cm}
\includegraphics[scale=0.5,trim=20 0 0 0]{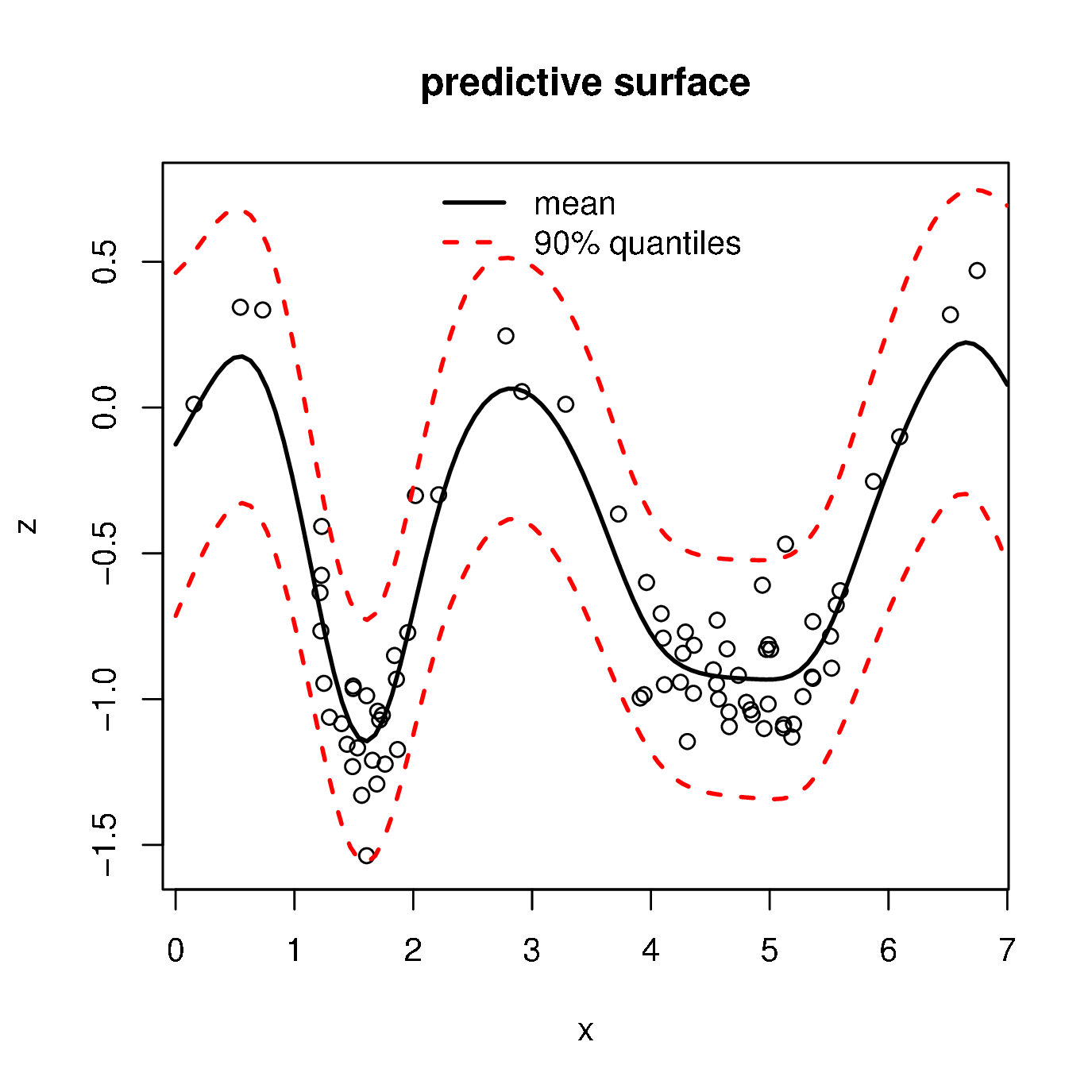}
\hspace{0.2cm}
\includegraphics[scale=0.5,]{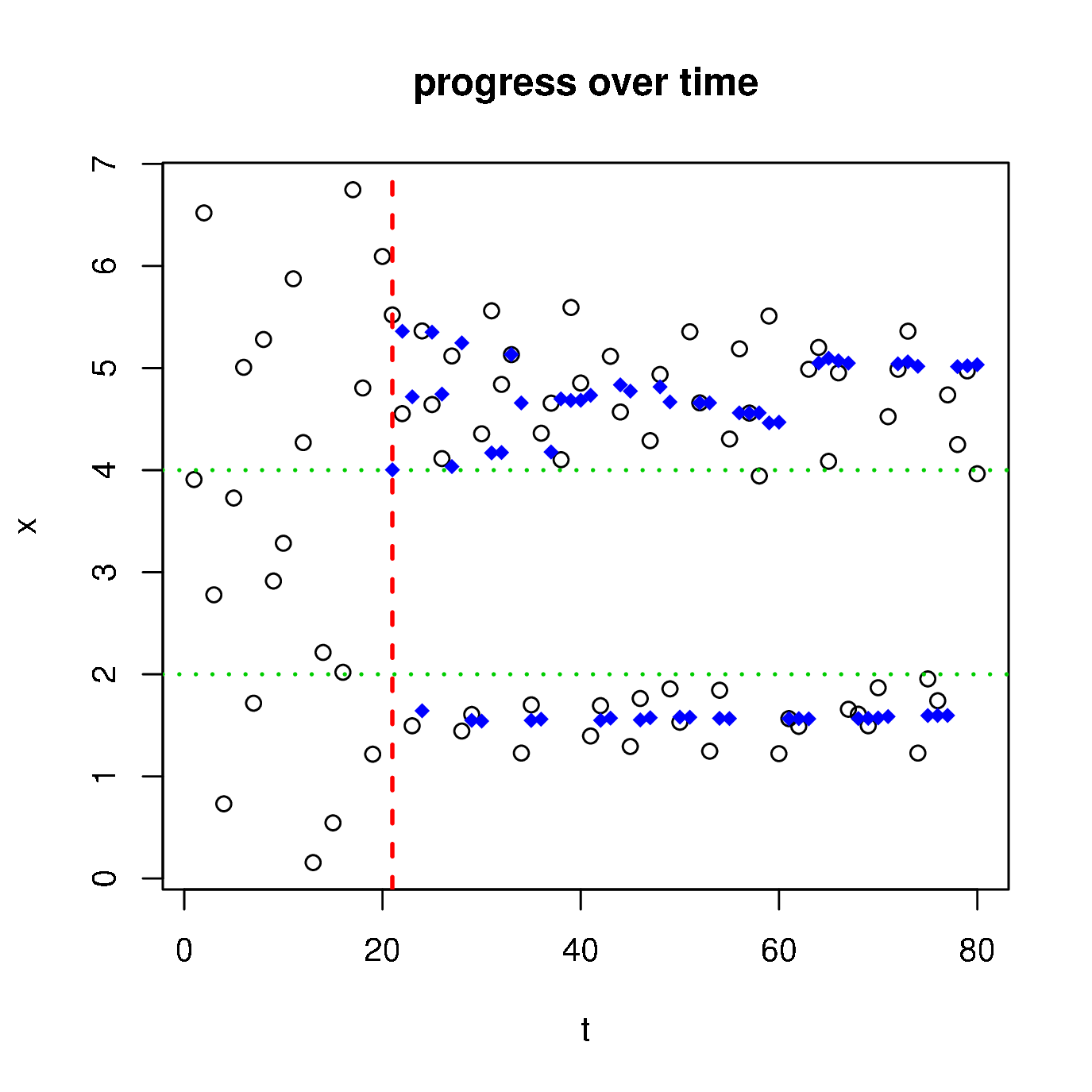}\\
\vspace{0.25cm}
\includegraphics[scale=0.5,trim=20 0 0 0]{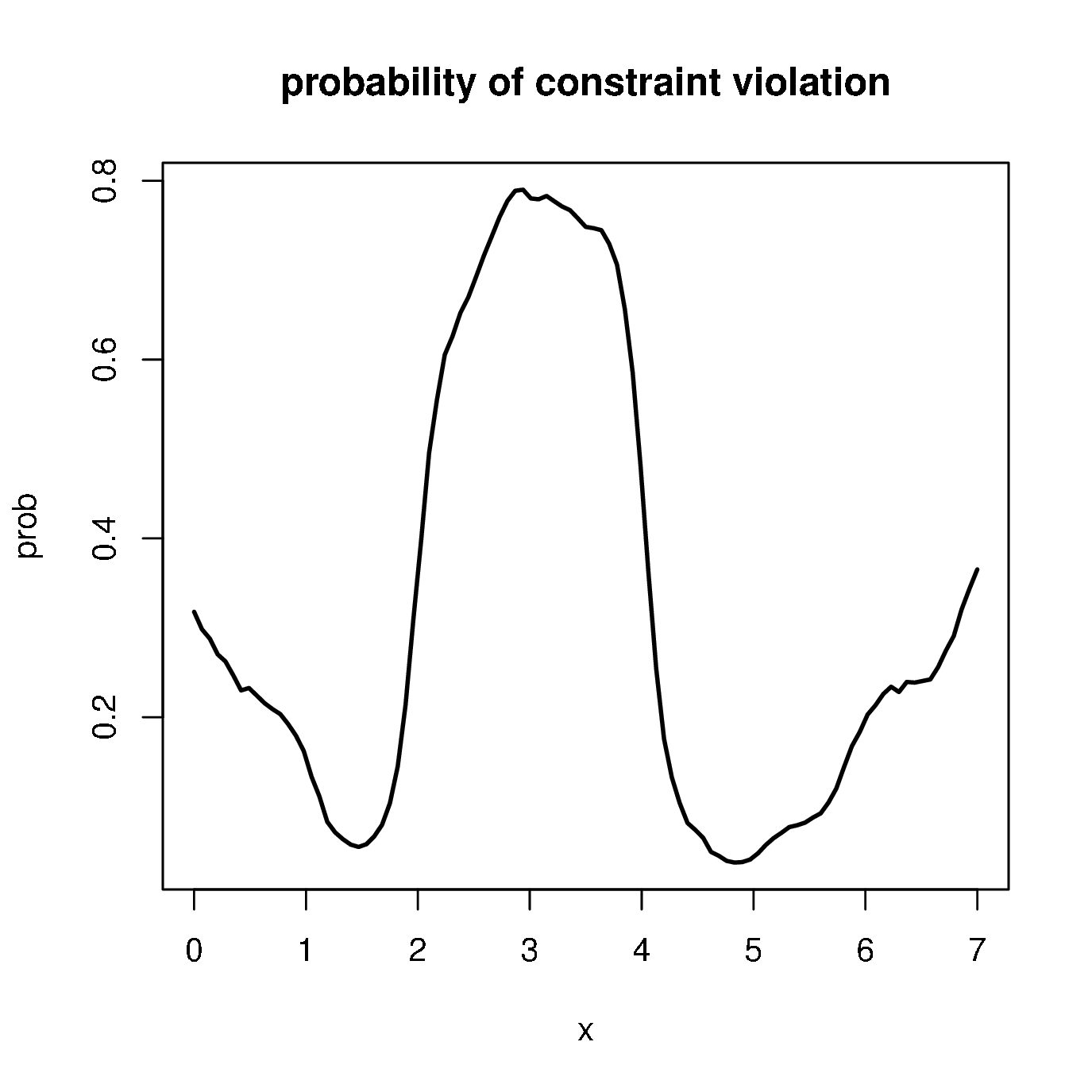}
\hspace{0.2cm}
\includegraphics[scale=0.5]{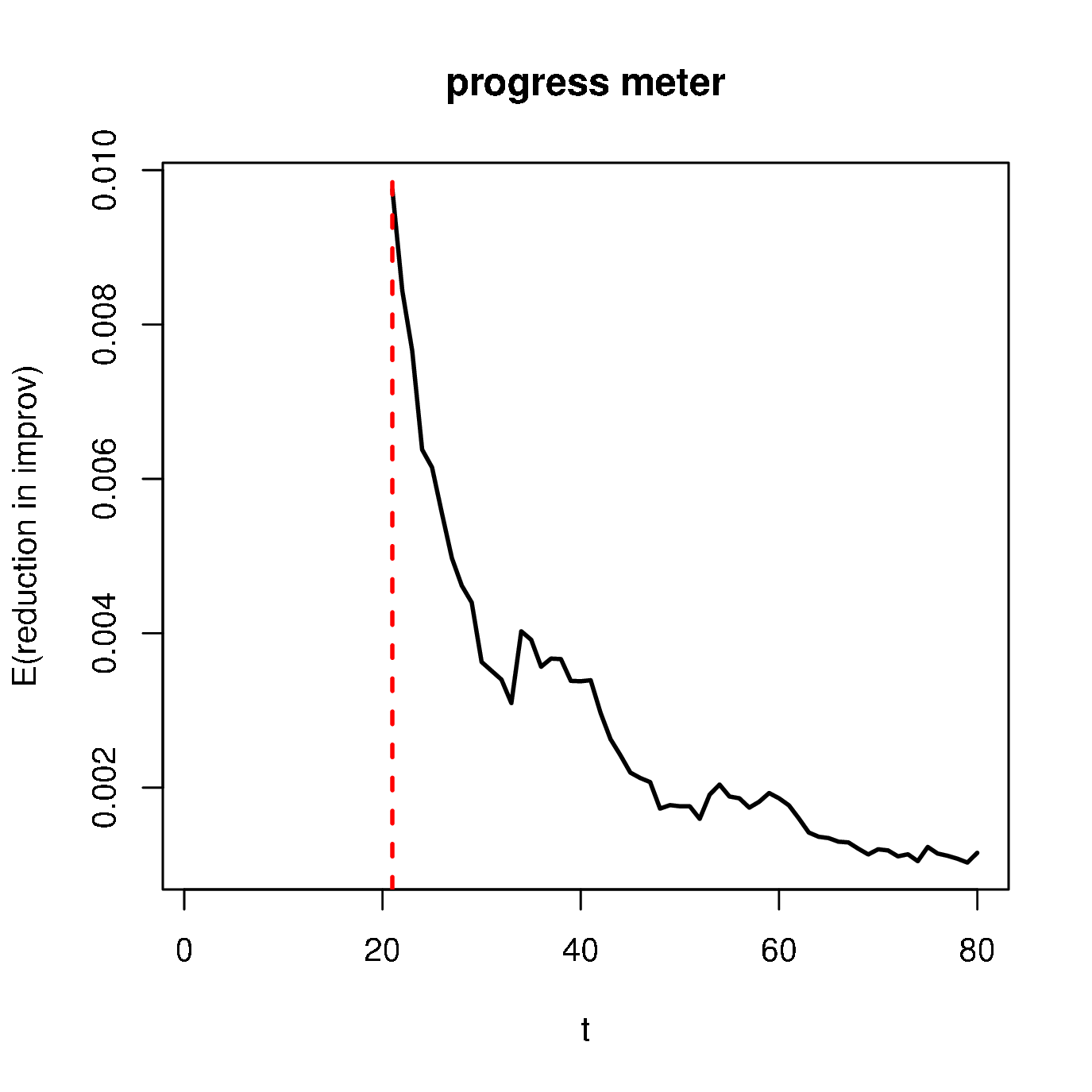}
\caption{Progress in 1-d optimization after 80 samples: {\em top-left}
  shows the posterior mean predictive surface (of $f_N$); {\em
    top-right} shows sampled $x$-values (open circles) and oracle
  candidates (closed) before and after the initial design, as
  separated by the vertical bar; horizontal lines indicate the unknown
  constraint region; {\em bottom-left} posterior mean of constraint
  (violation) surface ($c_N$); {\em bottom-right} the maximum of the
  log expected reduction in average improvement (\ref{eq:reduce}) over
  time.}
\label{f:taddy}
\end{figure}
Figure \ref{f:taddy} summarizes the results after the 80 total samples
were gathered.  Observe from the plots in the {\em top} row that most
samples (after the 20 initial ones) were gathered in the two local
minima, with a few taken outside $C$.  The oracle candidates (solid
circles) indicate the most likely locations of minima according to the
posterior predictive distribution.  The {\em bottom} panes show an
estimate of $c_N$ via the posterior mean probability of violating the
constraint ($\hat{P}(c_N(x) = 1)$), and a progress meter showing the
largest (log) expected reduction in average improvement
(\ref{eq:reduce}) at each round.  Observe how the ability to improve
upon the current minimum decreases over time, giving a heuristic
indication of convergence.

In our second example, the objective function for 2-d inputs $x =
(x_1, x_2)$ is given by
\begin{align}
f(x_1,x_2) & = -w(x_1)w(x_2), \;\;\;\;\; \mbox{where} \\
w(x) & = \exp\left(-(x-1)^2\right) + \exp\left(-0.8(x+1)^2\right) 
- 0.05\sin\left(8(x+0.1)\right) \nonumber
\end{align}
and observed without noise.  The constraint (satisfaction) region is
the interior of an ellipse defined by the 95\% contour of a bivariate
normal distribution centered at the origin, with correlation $-0.5$
and variance $0.75^2$.  The true global minimum is at $(x_1, x_2) =
(-1.408, -1.408)$, which does not satisfy the constraint.  There are,
however, three other local minima---two of which satisfy the
constraint.  The setup is as described above for the 1-d example
except that the optimization is initialized with 25 LHD samples, after
which 100 are gathered by IECI with 100 fresh candidates in each
round.
\begin{figure}[ht!]
\centering
\vspace{0.3cm}
\includegraphics[scale=0.5,trim=20 0 0 0]{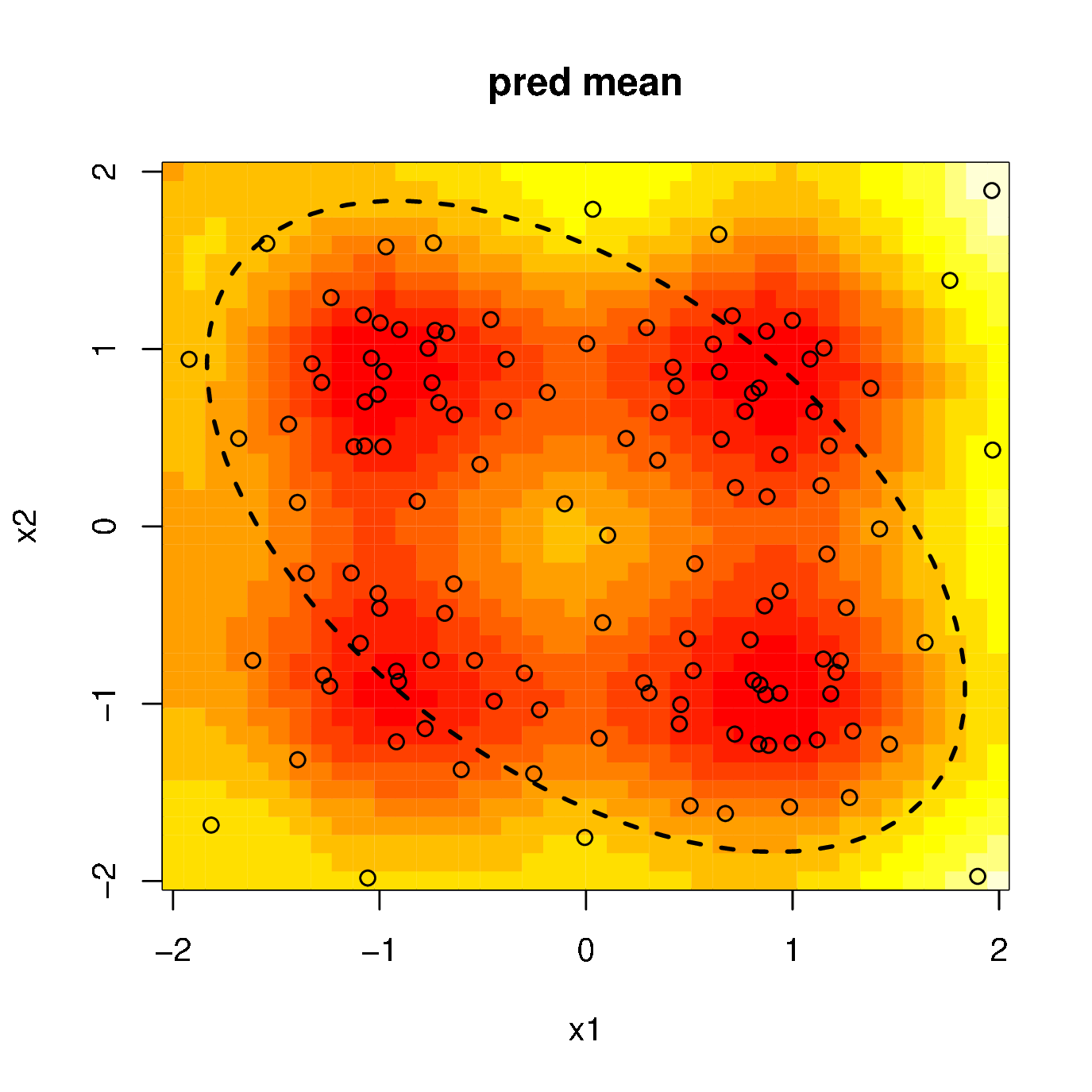}
\hspace{0.2cm}
\includegraphics[scale=0.5]{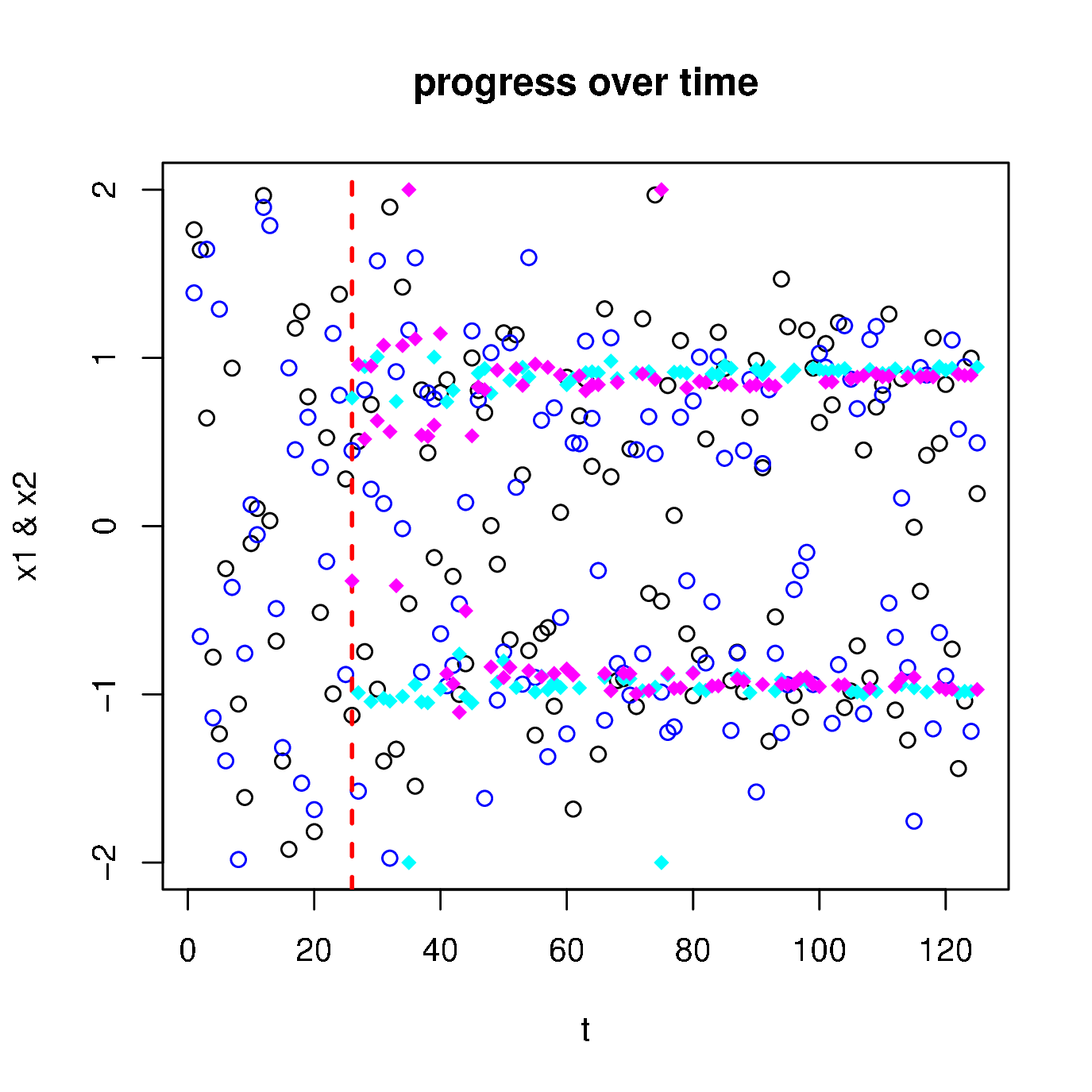}\\
\vspace{0.25cm}
\includegraphics[scale=0.5,trim=20 0 0 0]{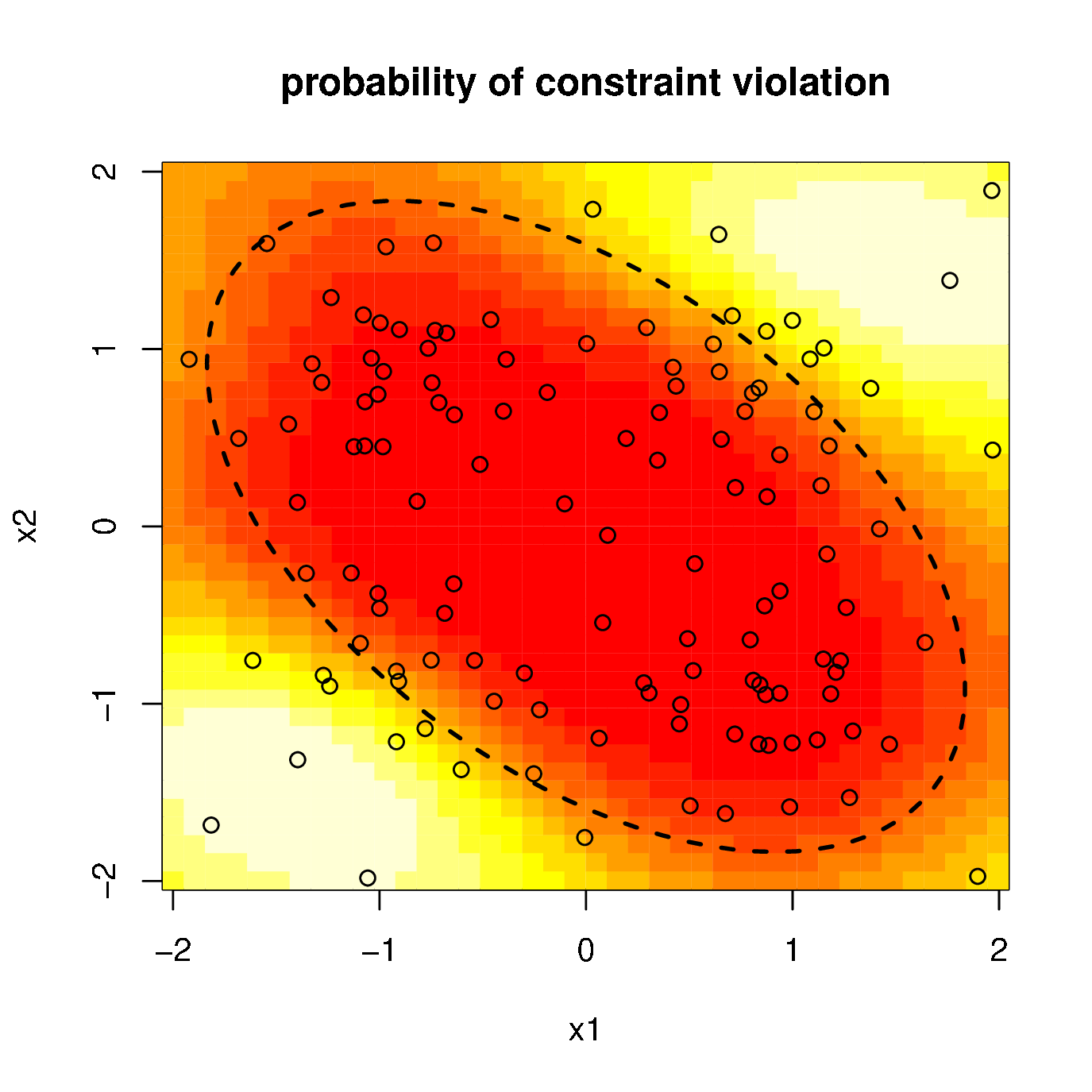}
\hspace{0.2cm}
\includegraphics[scale=0.5]{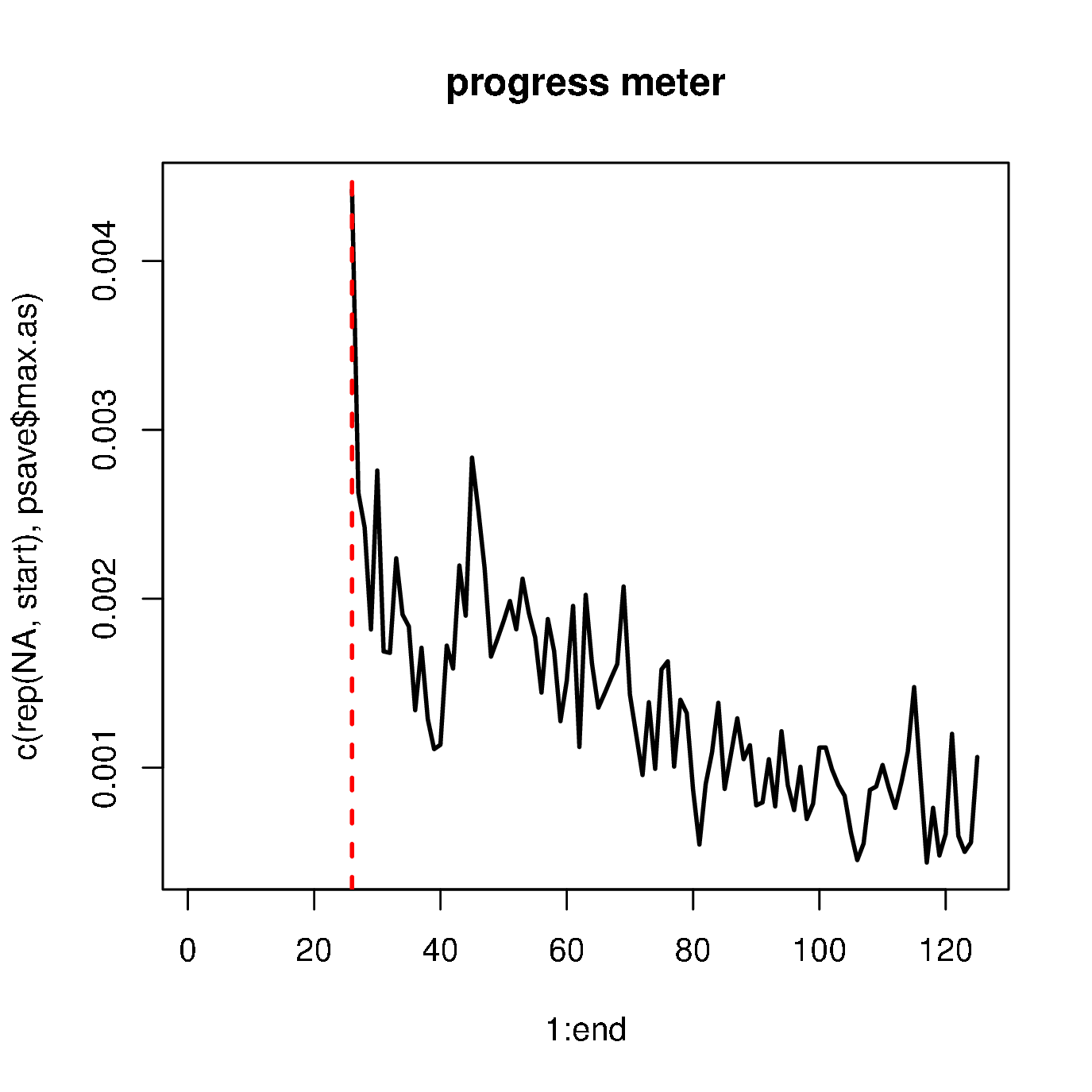}
\caption{Progress in 2-d optimization after 125 samples: {\em
    top-left}: posterior mean predictive surface; {\em top-right}:
  sampled $(x_1,x_2)$-values (open) and oracle candidates (closed);
  {\em bottom-left}: posterior mean of constraint surface; {\em
    bottom-right}: the progress meter (\ref{eq:reduce}).}
\label{f:nanprob}
\end{figure}
Figure \ref{f:nanprob} summarizes the results after the 125 total
samples were gathered.  Observe that very few samples were gathered
outside the unknown constraint region, except near the local minima.
It is sensible to sample heavily on the boundary of the constraint
region where the response is quickly changing and local minima are
likely to occur.  This is in case the global minimum is on the
boundary, and also helps to extract the GP parameters in regions of
highest importance.  Notice that large concentrations of samples occur
for the two minima well inside the constraint region.  But the {\em
  bottom-right} plot indicates that further progress can be made by
additional sampling.


\section{Health Policy Optimization}
\label{sec:rand}

Our motivating example involves a simulation of health care policy in
the United States.  The COMPARE simulator \citep{giro:2009} was
developed at the RAND Corporation to predict the effect of various
health care policies in terms of individual choices of health
insurance and the associated costs.  It is an agent-based
microsimulation model that uses a maximum utility approach to predict
the health insurance decisions of individuals, families, and firms as
a function of a wide range of inputs on available types of policies,
and on taxes, penalties, and regulations.  The population is simulated
based on Census Bureau data.  Additional datasets provide values for
many of the parameters in the simulation, and other parameters are set
as part of the possible policy interventions.  However, there are
several calibration parameters that are tuned so that when the
simulator is run on current policies, it makes predictions as close as
possible to the current observable situation in the United States.
Such a calibration can be viewed as a minimization problem, choosing
the values of the calibration parameters to minimize the discrepancy
between predictions and reality.  This setup is common for computer
simulators and has been investigated in the unconstrained setting
\citep[e.g.,][]{kennedy:ohagan:2001}.  What differs from the standard
setup here is the presence of unknown constraints.  

The simulator has a number of inputs and outputs, here we focus on a
subset deemed most important by our collaborators, the designers of
the simulator.  The inputs over which we optimize are a set of six
calibration parameters: utility tuning parameters for adults on ESI
programs, adults on individual programs, and adults on public
programs, and an analogous set of three parameters for children.  The
outputs of interest are the predicted counts in each type of insurance
(or the uninsured category) and the elasticities of response for the
key categories of adults in individual plans, adults in restricted
individual plans, uninsured adults, children in individual plans,
children in restricted individual plans, and uninsured children.  The
objective function specified by our collaborators is a combination of
the absolute errors in the predicted counts and the squares of the
predicted elasticities:
\[  
  Z({\bf x}) = \alpha_1 \sum_{j=1}^{4} |y_{aj} - \hat{y}_{aj}| +
  \alpha_2 \sum_{j=1}^{4} |y_{cj} - \hat{y}_{cj}| + 
  \sum_{k=1}^{4} \alpha_{3k} y_{ek}^2 \mathbb{I}_{\{ |y_{ek}|>1 \}} 
\]
where $\alpha_1$, $\alpha_2$, and $\alpha_{3k}$ are constants
specified by our collaborators that weight the pieces appropriately.
Our goal is to minimize this objective function under the constraint
that the elasticities for the insured are negative and the
elasticities for the uninsured are positive.  The elasticities can
only be found by running the simulator, so this set of constraints
fits under our unknown constraints regime.

\begin{figure}[htb]
\centering
\includegraphics[scale=0.64,angle=-90,trim=0 10 290 0]{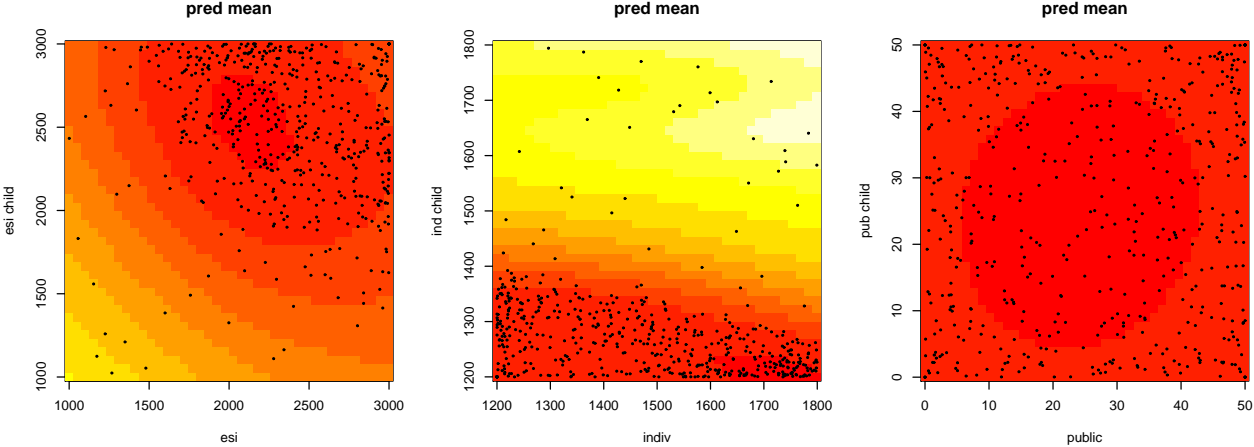}
\caption{Slices of the fitted response surface; dark shades are lower values.}
\label{f:comp.surf}
\end{figure}

Figure~\ref{f:comp.surf} shows pairwise slices of the fitted response
surface. The {\em left} panel shows how the fitted predicted surface
varies as a function of the parameters for adult and child ESI, when
the other four parameters are held fixed at a value around that which
produces the minimum response. The {\em middle} and {\em right} panels
vary by the parameters for individual programs and public programs
respectively. Dark shades are lower values, so it can be seen that
both ESI parameters need to be relatively high, the child individual
parameter needs to be low, and the other three parameters are
relatively less important.  The points plotted in the figure are the
550 total inputs sampled projected into the each of the three pairs of
input coordinates.

\begin{figure}[htb]
\centering
\includegraphics[scale=0.64,angle=-90,trim=0 10 290 0]{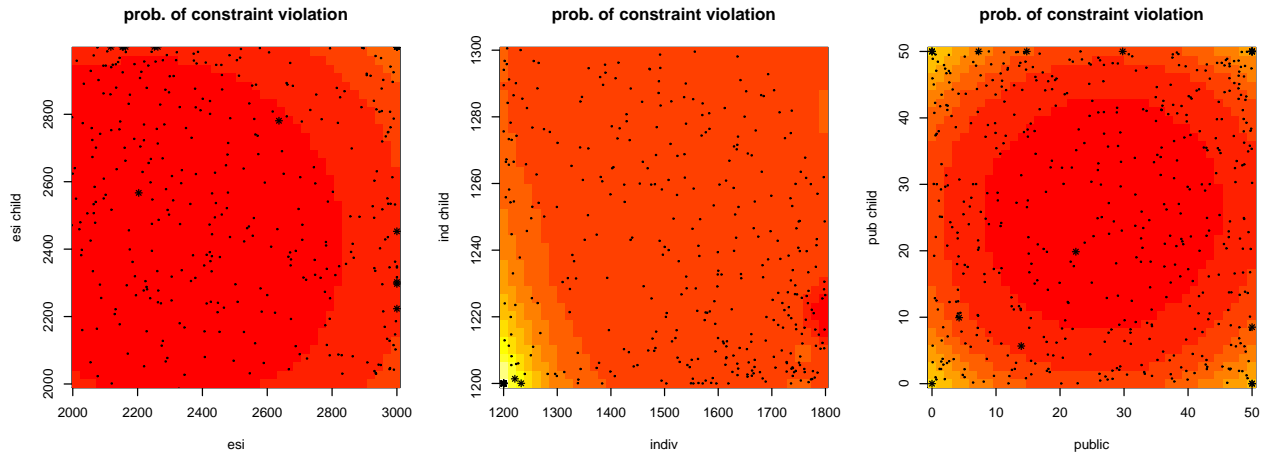}
\caption{Slices of the fitted probability of constraint violation;
  dark shades are lower values; sampled points violating the
  constraint are shown with asterisks.}
\label{f:comp.prob}
\end{figure}

Figure~\ref{f:comp.prob} shows the fitted probability of a constraint
violation over the portions of the space which were routinely sampled.
As seen in Figure~\ref{f:comp.surf}, some regions are not well-sampled
because they do not help in finding the minimum, the goal of the
problem.  These sparsely sampled regions do not provide much
information for estimating the probability of a constraint violation
(which is not the primary goal of the problem), and so the estimated
values are overly influenced by the prior mean.  Thus we only display
parts of the regions in the first two plots to better show the
estimated probabilities.  Sampled points which violated the
constraints are shown with asterisks.  One can see that the largest
probabilities of constraint violations occurred for large values of
the ESI parameter, for jointly small values of the individual and
child individual parameters, and for values of the public and child
public parameters which are in the corners of the space.

\begin{figure}[htb]
\centering
\vspace{0.2cm}
\hspace{0.2cm}
\includegraphics[scale=0.6]{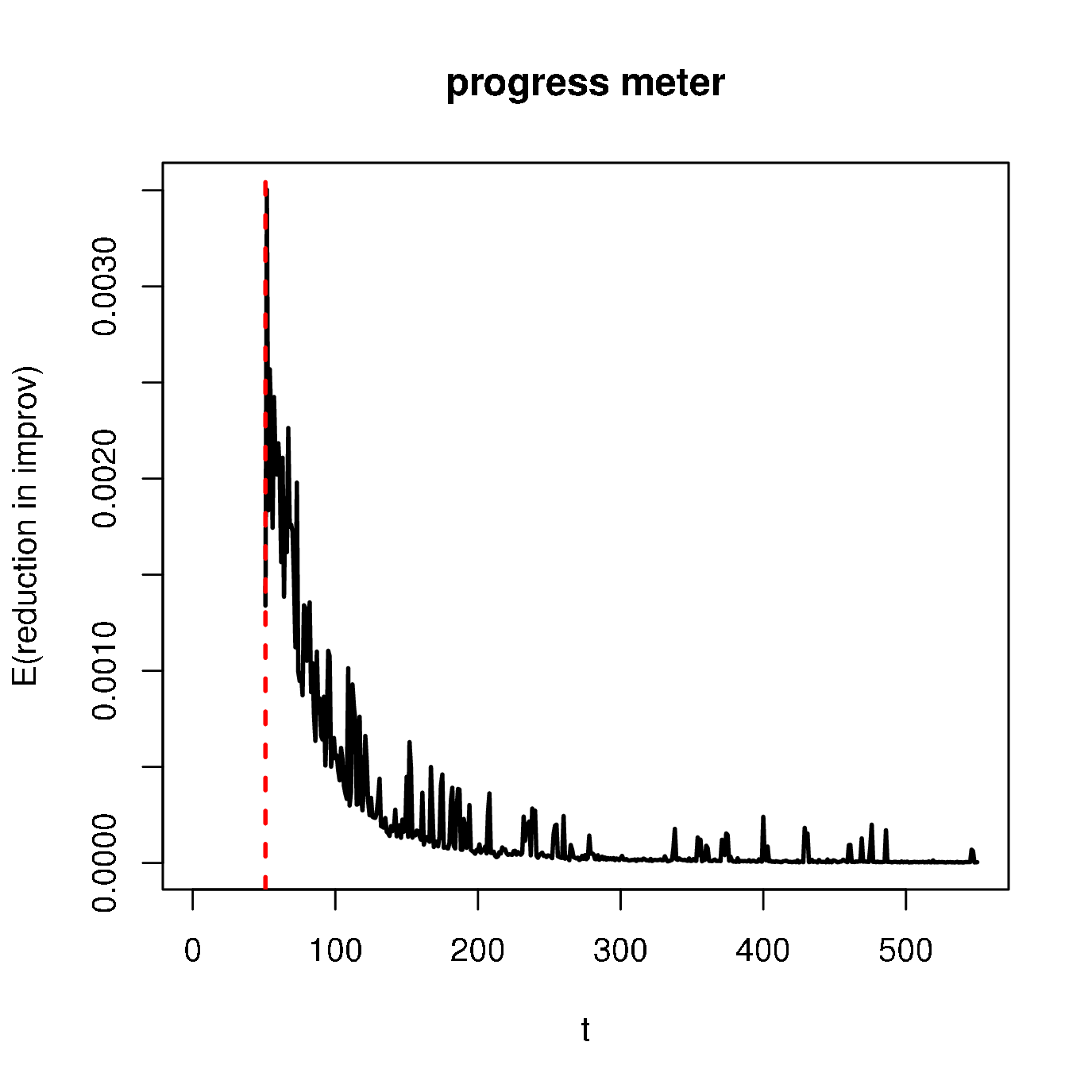}
\caption{Progress meter (\ref{eq:reduce}) for the health policy
  optimization.}
\label{f:prog}
\end{figure}

Figure \ref{f:prog} shows the progress meter (\ref{eq:reduce}) over
the 500 optimization rounds which can be used as a heuristic check of
convergence.  As in previous examples, the noisiness in the meter is
due to the LHD predictive grid of 100 candidates at which the IECI is
evaluated in each round.  After about 250 samples the IECI seems to
have ``bottomed-out''.  However, further progress can be made to
reduce the frequency and magnitude of the ``up-spikes'' in the
remaining optimization rounds, and thereby obtain higher confidence
that the constrained global minimum has been obtained.

\section{Discussion}
\label{sec:discuss}

We have introduced a statistical approach to optimization under
unknown constraints by an integrated conditional expected improvement
(IECI) statistic. The idea is to consider how the improvement at
reference locations ($y$) conditional on candidates ($x$) may be used
to augment a design.  Without considering constraints, the resulting
statistic is a less greedy---aggregated---version of the standard
expected improvement (EI) statistic. Another way to obtain a less
greedy EI is to raise the improvement to a power $g$
\citep{scho:welc:jone:1998}. The IECI approach, by contrast, does not
require such a tuning parameter. In the presence of unknown
constraints, IECI allows us to coherently consider how design
candidates adjust the improvement at reference locations believed to
satisfy the constraint. Our method was illustrated on two synthetic
examples and a motivating problem from health care policy.  An
implementation is provided in the {\tt plgp} package on CRAN.

We envisage many ways that our methodology may be extended and
improved. Understanding of convergence of statistical optimization
algorithms is scant at best, and IECI is no exception. While we
provide a sensible heuristic that seems to work well in our examples,
much remains to be done in this area. It may also be sensible to model
the constraint as a function of the inputs ($x$) and the real-valued
response ($Z(x)$). An example of where this would be handy is when $C
= \{ x: Z(x) < k\}$, for some constant $c$. Our dual-GP modeling
framework may easily be extended to allow uncertainty in $Z$
(real-valued) responses to filter through, as predictors, into the
surrogate model for the classification labels.  A more difficult
extension involves accommodating {\em hidden constraints}
\citep{lee:gra:link:gray:2010}: where evaluation of the real-valued
response fails, e.g., due to a lack of convergence in a simulation.
Finally, it may be worthwhile to consider surrogate models beyond GPs.
Dynamic trees for regression and classification show considerable
promise \citep{tad:gra:pols:2010}.

\subsection*{Acknowledgments}

This research was initiated at a workshop at the American Institute of
Mathematics (AIM) on Derivative-Free Hybrid Optimization Methods for
Solving Simulation-Based Problems in Hydrology, and was also partially
supported by NSF grant DMS-0906720 to HKHL and EPSRC grant
EP/D065704/1 to RBG.  The authors would like thank Crystal Linkletter
for interesting discussions at AIM, the RAND Corporation, Health
division, for the use of COMPARE, and Federico Girosi, Amado Cordova,
and Jeffrey Sullivan in particular for their help with the simulator.

\bibliography{intei}
\bibliographystyle{jasa}

\end{document}